\documentclass[12pt]{iopart}
\usepackage{graphicx}
\begin{document}

\title[Cosmology with type-Ia supernovae]{Cosmology with type-Ia supernovae}

\author{Ramon Miquel}

\address{Lawrence Berkeley National Laboratory (LBNL) \\
1 Cyclotron Rd, Berkeley, CA 94720, USA\\
{\em{and}} \\
Instituci\'o Catalana de Recerca i Estudis Avan\c{c}ats (ICREA) \\
Institut de F\'{\i}sica d'Altes Energies (IFAE) \\
Edifici Cn, Campus UAB, E-08193 Bellaterra (Barcelona), Spain}
\ead{ramon.miquel@ifae.es}
\begin{abstract}
I review the use of type-Ia supernovae (SNe) for
cosmological studies. After briefly recalling the main features of type-Ia SNe
that lead to their use as cosmological probes, 
I briefly describe current and planned type-Ia SNe surveys, with
special emphasis on their physics reach in the presence of systematic
uncertainties, which will be dominant in nearly all cases.
\end{abstract}
\vspace{2pc}
\section{Introduction}\label{sec:intro}
Over the last ten years type-Ia supernovae (SNe) have been established as a prime cosmological 
tool.
In 1998, the study of the redshift-luminosity relation (Hubble diagram) for near-by and distant 
supernovae~\cite{perlmutter99,riess98}
provided the ``smoking gun'' for the accelerated expansion of the universe and the existence of
the mechanism that drives it, code-named ``dark energy''. Since then, several surveys have added
substantial statistics to the Hubble diagram~\cite{astier06} and extended it to higher 
redshifts~\cite{knop03,riess04}.

The goal of the near-future type-Ia supernova surveys is to help determine the
properties of the dark energy component of the universe, as encoded in its equation of
state parameter $w \equiv p/\rho$, where $p$ is its pressure and $\rho$ its energy density.
The equation of state parameter is customarily parameterized
as~\cite{french,linder01} $w(z) = w_0 + w_a\left(1-a\right)$, where $w_0$ is the equation
of state parameter now (which has to fulfill $w_0 < -1/3$ in order to drive the current
accelerated expansion of the universe), and $w_a = -\rmd w / \rmd \ln a|_0$ is a measure of 
the current rate of change of $w$ with time. 
Here $z$ is the redshift and $a=\left(1+z\right)^{-1}$ 
is the expansion parameter of the universe, with the current value being $a_0 = 1$, 
corresponding to $z=0$. For a cosmological constant, we have $w_0 = -1$, $w_a = 0$.
A first goal is to determine $w_0$ and $w_a$ with enough accuracy to establish whether the dark 
energy is ``just'' a cosmological constant or it has a dynamical origin.

The reach in $w_0$ and $w_a$ in current and future high-statistics type-Ia SNe 
surveys is already limited by systematic uncertainties. A lot of effort is being put in
gathering well-measured (both photometrically and spectroscopically) samples of 
near-by SNe~\cite{SNF,SDSS}
in order to study their properties in detail and constrain the systematic uncertainties. In
designing new surveys it is of the utmost importance to pay attention to systematics. Predictions
based solely on statistical reach are doomed to be proved over-optimistic and misleading when
data arrive.

The outline of this note is as follows. In section~\ref{sec:sne-cosmo} we will briefly present
the main astrophysical and observational features of type-Ia SNe, and will introduce the Hubble
diagram from the Friedman equations. 
Systematic uncertainties are discussed in some detail in section~\ref{sec:syst}. 
The current state-of-the-art ground-based
SuperNova Legacy Survey (SNLS) is discussed is section~\ref{sec:snls}, 
while the planned SuperNova
Acceleration Probe (SNAP) mission is presented in section~\ref{sec:snap}. In both cases, emphasis
is given to their limiting systematic uncertainties. Finally, we summarize the note in
section~\ref{sec:summary}.
\section{Type-Ia supernovae as cosmological tools}\label{sec:sne-cosmo}
\subsection{Type-Ia supernovae}\label{sec:sne}
Observationally, type-Ia supernovae are defined as supernovae without any hydrogen lines in
their spectrum, but with a prominent, broad silicon absorption line (Si-II) at about 400~nm in
the supernova rest frame. The progenitor is understood to be a binary system in which a white
dwarf (no hydrogen) accreets material from a companion star (possibly another white dwarf). The
process continues until the mass of the white dwarf approaches the Chandrasekhar limit,
at which point a thermonuclear runaway explosion is triggered. 

The fact that all type-Ia SNe have a similar mass~\footnote{See~\cite{2003bb} for an intriguing
exception.} helps explain their remarkably homogeneity. Type-Ia SNe are very
homogeneous in luminosity, color, spectrum at maximum light, etc. Only small and correlated 
variations of these quantities are observed. They are very bright events with absolute magnitude 
in the $B$ band reaching $M_B \sim -19.5$ at maximum light. The raise time and decay time of their
light curve (magnitude as a function of time) are, respectively, 15--20 days and $\sim$ 2 months,
in the SN rest frame.

\begin{figure}[tb]
\begin{center}
\includegraphics[scale=0.85]{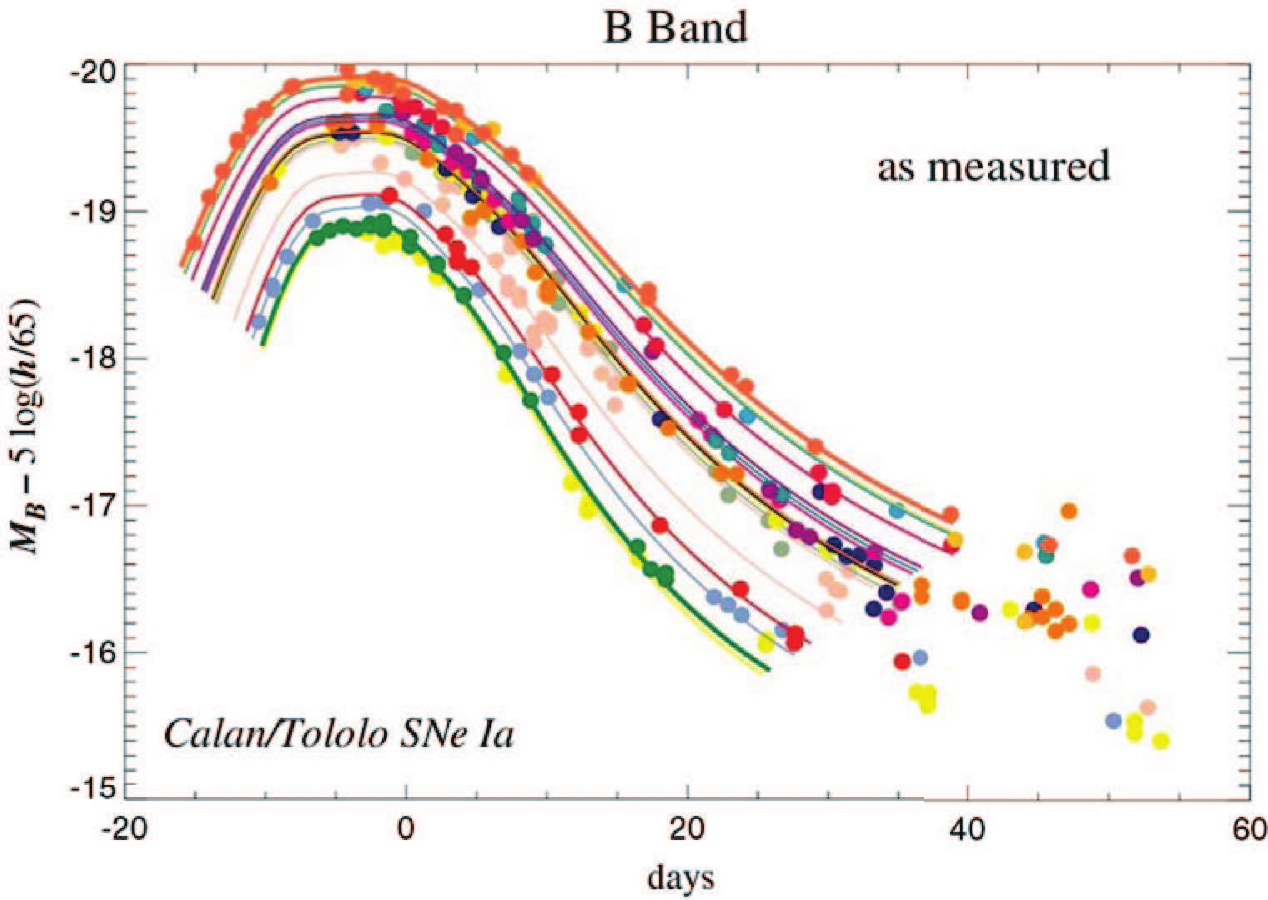}
\caption{\small B-band light curves of the Cal\'an/Tololo type-Ia supernova sample before any
duration-magnitude correction.
\label{fig:Calan-Tololo-no-stretch}}
\end{center}
\end{figure}

\begin{figure}[tb]
\begin{center}
\includegraphics[scale=0.85]{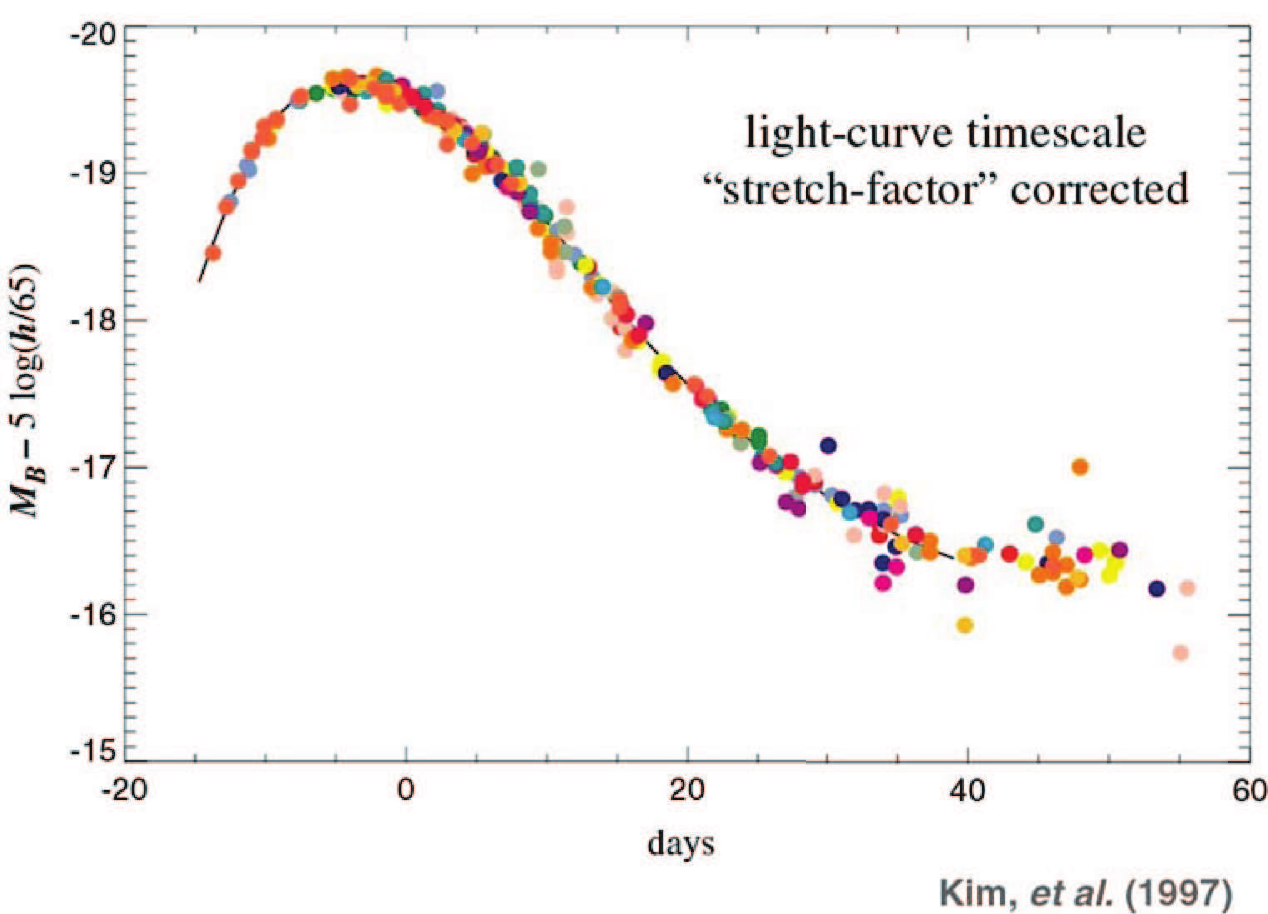}
\caption{\small Same light curves of fig.~\ref{fig:Calan-Tololo-no-stretch} after applying the
``stretch'' duration-magnitude correction of ref.~\cite{stretch}.
\label{fig:Calan-Tololo-stretch}}
\end{center}
\end{figure}

In 1992 Mark Phillips~\cite{phillips92} found that for near-by SNe there was a clear correlation
between their intrinsic brightness at maximum light and the duration of their light curve,
so that brighter SNe last longer (see fig.~\ref{fig:Calan-Tololo-no-stretch}). 
Several empirical
techniques~\cite{stretch,Dm15,mlcs,batm} have been developed since then to make use to this
correlation to turn type-Ia SNe into standard candles, with a dispersion on their peak magnitude
of only 0.10--0.15~mag, corresponding to a precision of about 5--7\% in distance, and,
therefore, in lookback time to the explosion. Figure~\ref{fig:Calan-Tololo-stretch} shows
the same SNe light curves of fig.~\ref{fig:Calan-Tololo-no-stretch} after applying
the ``stretch'' technique of~\cite{stretch}, so called because it basically amounts to a simple 
stretching of the time axis, showing the good uniformity achieved.

\begin{figure}[tb]
\begin{center}
\includegraphics[scale=0.85]{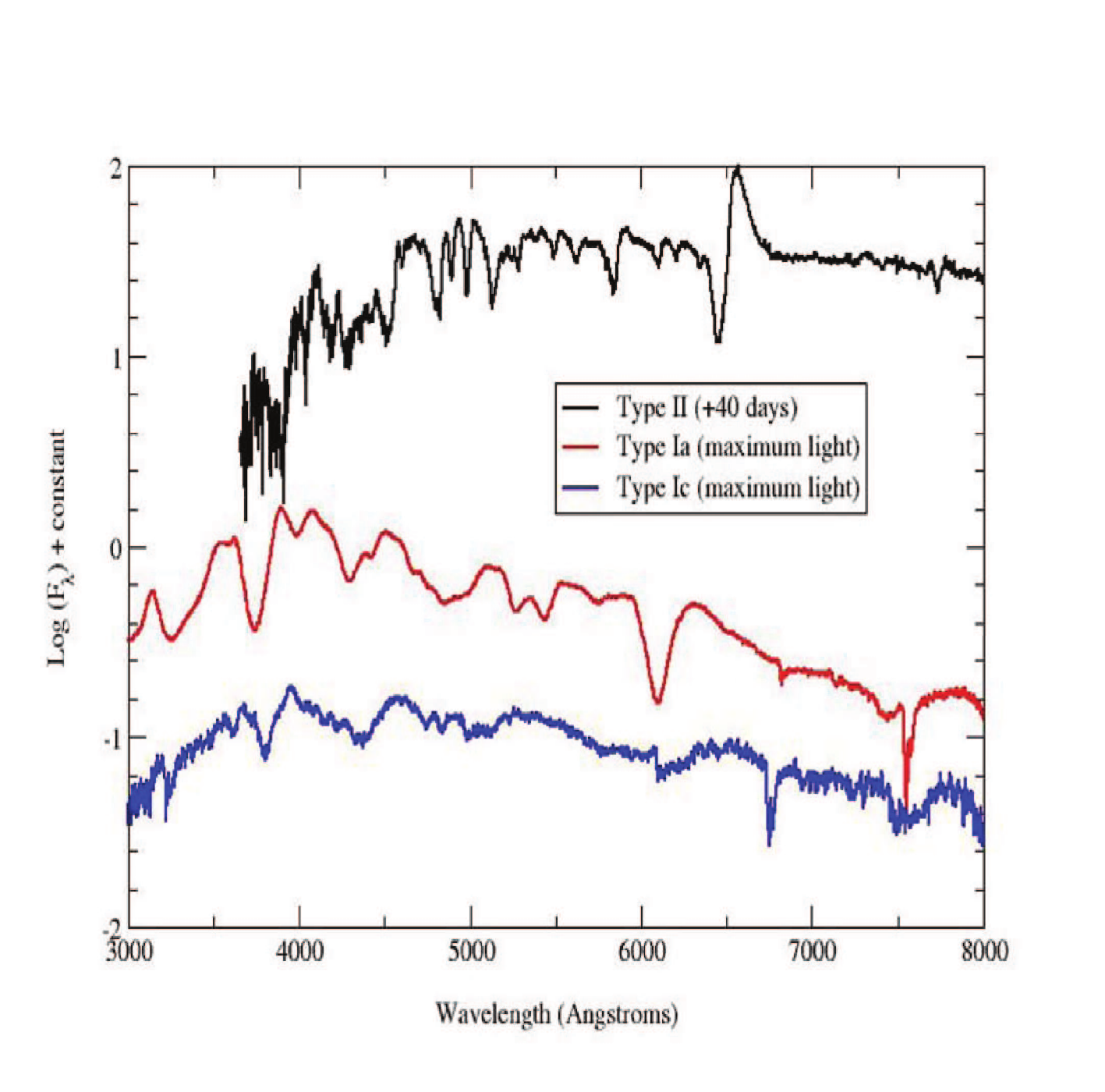}
\caption{\small Measured spectra of three supernovae. From top to bottom: type II, type Ia, 
type Ic. The Si-II feature identifying a type-Ia SNe is clearly visible in the middle spectrum
at about 600~nm.
\label{fig:spectrum}}
\end{center}
\end{figure}

While light-curves are determined with photometric measurements in several broadband filters,
spectroscopy near maximum light serves the dual purpose of unambiguously identifying the object
as a type-Ia SN and at the same time determining its redshift. Both goals are achieved by
comparing the measured spectrum to templated spectra from well-measured near-by
type-Ia supernovae. 
The key feature of the spectrum is the Si-II absorption line, whose detection identifies the SN
as a type Ia, and whose position determines the redshift. Figure~\ref{fig:spectrum} shows 
spectra of three supernovae: from top to bottom, a type II, a type Ia, and a type Ic . 
The Si-II feature can be seen clearly in the type-Ia spectrum.

\subsection{Cosmology}\label{sec:cosmo}
Once the stretch-corrected magnitude and redshift are determined, the supernova can be put into 
a Hubble diagram in order to measure the cosmological parameters. The Hubble diagram is a plot
of measured magnitude vs. redshift. Since the apparent magnitude of a standard candle gives us
its distance and the time $t$
at which the light was emitted, and the redshift gives the cosmic 
expansion parameter $a(t)$, a Hubble diagram populated with SNe at different distances gives us the
history of the expansion of the universe. Since the expansion rate of the universe is determined
by its matter-energy content, it is clear that type-Ia SNe can tell us about the properties
of the contents of the universe, and, in particular, of the dark energy component.

Assuming that the universe is homogeneous and isotropic at large scales leads to the 
Friedman-Lema\^{\i}tre-Robertson-Walker (FLRW) universe defined by the metric
$\rmd s^2 = \rmd t^2 - a^2(t)(\rmd r^2/(1-kr^2) 
+ r^2 (\rmd\theta^2 + \sin^2\theta \rmd\phi^2))$, 
where $t$ is the proper time and
$(r,\theta,\phi)$ are co-moving coordinates. For a flat universe, that we will assume in most
of the following,
$k=0$. For the FLRW metric, Einstein's field equations of general relativity reduce to the
so-called Friedman-Lema\^{\i}tre equations:
\begin{eqnarray}
\frac{\ddot{a}}{a} = -\frac{4\pi G}{3}\left(\rho+3p\right) \label{eq:friedman1} \\
\left(\frac{\dot{a}}{a}\right)^2 = \frac{8\pi G}{3} \rho -\frac{k}{a^2} \label{eq:friedman2}\ .
\end{eqnarray}
From the first equation, it is clear that in order for the expansion of the universe to 
accelerate ($\ddot{a}>0$), it is necessary that $\rho + 3p < 0$, or $w < -1/3$.

Since both $\rho$ and $p$ evolve with time, in order to solve for $a(t)$ we need an extra 
equation. This can be the equation of state for each component of the universe, relating
its energy density with its pressure. For matter (ordinary or dark), $p=0$, so $w=0$.
For radiation, we have the relativistic gas relationship $p=\rho/3$, so $w=1/3$. As mentioned
before, for the cosmological constant one has $p=-\rho$, or $w=-1$. Assuming a flat universe,
Eqs.~(\ref{eq:friedman1},\ref{eq:friedman2}) can be used to obtain the relationship
\begin{equation}
\frac{\rmd\rho}{\rmd a} = -3(1+w)\frac{\rho}{a} \ ,
\end{equation}
from which, assuming a constant equation of state $w$, one gets:
\begin{equation}
\rho = \rho_0 a^{-3(1+w)} \ ,
\end{equation}
which results in $\rho = \rho_0 a^{-3} = \rho_0 (1+z)^3$ for matter, 
$\rho = \rho_0 a^{-4} = \rho_0 (1+z)^4$ for radiation, and $\rho = \rho_0$ for a cosmological
constant. Introducing the Hubble parameter $H=\dot{a}/a$ and defining the critical density as 
$\rho_c = 3H_0/8\pi G$, where $H_0$ is the Hubble parameter now, we can cast 
eq.~(\ref{eq:friedman2}) as
\begin{equation}
\label{eq:friedman3}
H^2 = H^2_0 \left[\Omega_M(1+z)^3 + \Omega_r(1+z)^4 + \Omega_{DE}(1+z)^{3(1+w_{DE})}\right] \ ,
\end{equation}
where we have introduced the current normalized densities $\Omega_i \equiv \rho^i_0 / \rho_c$,
for $i=M$ (matter), $r$ (radiation) and $DE$ (dark energy). The term proportional to
$\Omega_r$ can be safely neglected for all purposes, 
at least for moderate values of $z$ ($z < 5000$).
It is clear from this equation that
by measuring $H$ at different times (the history of the expansion of the universe as
provided by type-Ia SNe), one can learn about the properties of the constituents of
the universe, $\Omega_M, \Omega_{DE}, w_{DE}$, etc.
\subsection{The Hubble diagram}\label{hubble}
\begin{figure}[tb]
\begin{center}
\includegraphics[scale=0.65]{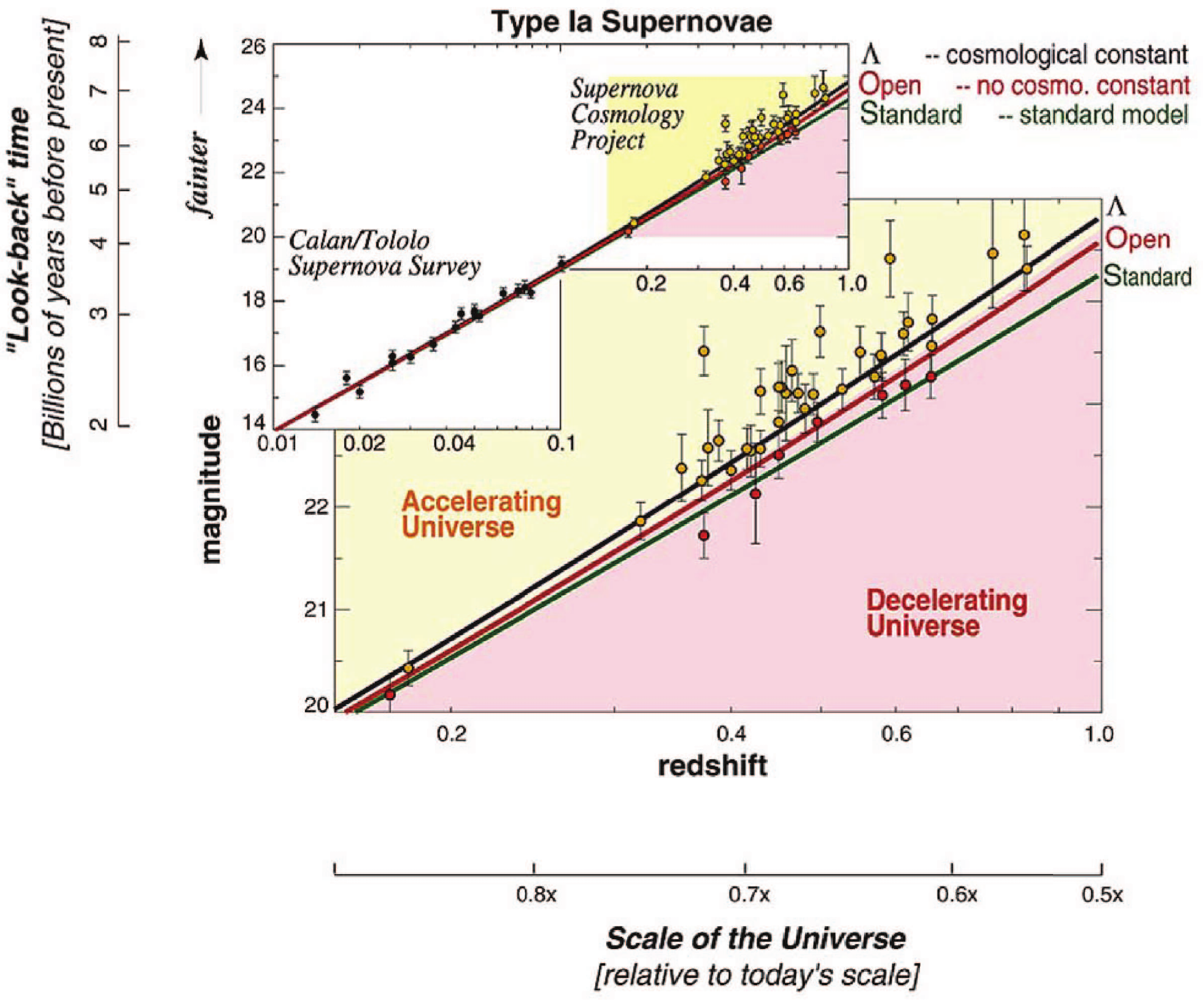}
\caption{\small Example of apparent magnitude vs. redshift Hubble diagram, from the 1998 
Supernova Cosmology Project results~\cite{perlmutter99}.
\label{fig:hubble}}
\end{center}
\end{figure}

Standard candles (or, in the case of type-Ia SNe, ``standardizable'' candles) provide a
measurement of the luminosity distance $d_L$ as a function of redshift. $d_L$ can be defined
through the relation $\phi = \frac{L}{4\pi d_L^2}$, where $L$ is the intrinsic luminosity,
and $\phi$ the flux, so that $d_L$ is the ``equivalent distance'' in a Euclidean, non-expanding
universe. It is easy to see that $d_L(z) = (1+z)\, r(z)$, where $r(z)$ is the co-moving
distance to the source at redshift $z$. Recalling that light travels in
geodesics ($\rmd s^2 = 0$), we can easily compute $r(z)$ from the FLRW metric as
\begin{equation}
\label{eq:comoving}
r(z) = \int_1^2 \rmd r = \int_1^2 \frac{\rmd t}{a} = \int_1^2 \frac{\rmd a}{a\dot{a}} = 
\int_0^z \frac{\rmd z'}{H(z')} \ ,
\end{equation}
where for simplicity we have assumed a flat universe.
Astronomers measure fluxes as apparent magnitudes:
\begin{eqnarray}
\label{eq:hubble}
m(z) \equiv -2.5\log\left(\phi/\phi_0\right) = {\cal M} + 5\log\left[H_0 d_L(z)\right] \\
{\cal M} \equiv M + 25 - 5\log\left[H_0/100\,\mathrm{kms^{-1}Mpc^{-1}}\right] \, , \nonumber
\end{eqnarray}
where $M$ is the (assumed unknown) absolute magnitude of a type-Ia SN, related to $-2.5\log L$.
The flux $\phi_0$ defines the zero point of the magnitude system used.
It should become clear from eqs.~(\ref{eq:comoving}) and (\ref{eq:friedman3}) that, contrary to
the appearances, eq.~(\ref{eq:hubble}) does not depend on $H_0$. An example of a Hubble
diagram can be seen in fig.~\ref{fig:hubble}. 
By measuring apparent magnitudes and redshifts
from a set of type-Ia supernovae, one can measure different integrals of $H_0/H(z)$, which
according to eq.~(\ref{eq:friedman3}) are sensitive to the cosmological parameters. Note that
in the standard cosmological analyses
${\cal M}$ is considered a nuisance parameter and it is determined simultaneously from the data.
%
%
%
\section{Systematic uncertainties}\label{sec:syst}
The statistical uncertainties in the Hubble diagram are dominated by the intrinsic supernova
peak magnitude dispersion $\sigma_{\mathrm int} = 0.10-0.15$. Since this error is uncorrelated
from supernova to supernova, in a redshift bin with ${\cal O}(100)$ SNe (a quantity most current
and all near-future surveys will achieve), the statistical error will be 
$\sigma_{\mathrm stat} = 0.01-0.02$. Since many systematic uncertainties are expected to be
fully correlated for SNe at similar redshifts, but uncorrelated otherwise, it is clear that
systematic errors of order a few per cent will be important, and, in many cases, already dominant.

A comprehensive study of systematic errors affecting type-Ia SNe distance measurements can be
found in~\cite{KLMM}. We will only cover the more relevant ones in the following.
\subsection{Host galaxy dust extinction}\label{sec:dust}
Dust in the path between the supernova and the telescope attenuates the amount of light measured.
Milky Way dust is well measured and understood~\cite{schlegel}, while intergalactic dust has a
negligible effect. In contrast, dust in the supernova host galaxy can lead to a 
substantial dimming of the SN
light. Ordinary dust absorbs predominantly in the blue, leading to a reddening of the SN
colors. The amount of reddening can be measured, and from it, the amount of extinction can be
determined, provided the extinction law (extinction as a function of wavelength) is known.
The usual extinction law~\cite{CCM} reads:
\begin{equation}
m_j \to m_j + A_V\left(a(\lambda_j) + \frac{b(\lambda_j)}{R_V} \right) 
= m_j + E(B-V) \left(R_V a(\lambda_j) + b(\lambda_j)\right) \, ,
\end{equation}
where $E(B-V)$ is the excess $B-V$ color over the expected one, $R_V\approx 3.1$ in 
near-by galaxies is sometimes called the extinction law, $A_V = R_V E(B-V)$ 
is the increase in magnitude in the $V$ band due to dust, 
$a(\lambda)$ and $b(\lambda)$ are known functions, with 
$a(\lambda_V) = 1$, $b(\lambda_V) = 0$, and all wavelengths are in the SN rest frame. 
In order to correct $m_j$ we need to know
$E(B-V)$ and $R_V$. The former can be determined from photometry in at least two bands. The
latter is more complicated. Although it can in principle be measured directly from three-band
photometry, in practice, the lever-arm is limited. Furthermore, current surveys do not
have precision photometry in three bands for all their SNe. Several alternative approaches
have been used in the literature. Riess et al.~\cite{riess04} assume $R_V = 3.1$ everywhere;
Astier et al.~\cite{astier06} instead determine one single effective $R_V$ for all their
distant SNe, finding a much lower value $R_V = 0.57 \pm 0.15$. However, this parameter 
effectively includes any other effect that might correlate SN color and magnitude. The proposed
SNAP satellite mission~\cite{snap} with its nine filters will determine $R_V$ for each SN
independently, since it will have precision optical and near-infrared photometry for all
their SNe in at least three and up to nine bands. Clearly, given the uncertainties on the
value of $R_V$ in distant galaxies, this looks like the most conservative approach.

\begin{figure}[tb]
\begin{center}
\includegraphics[scale=0.75]{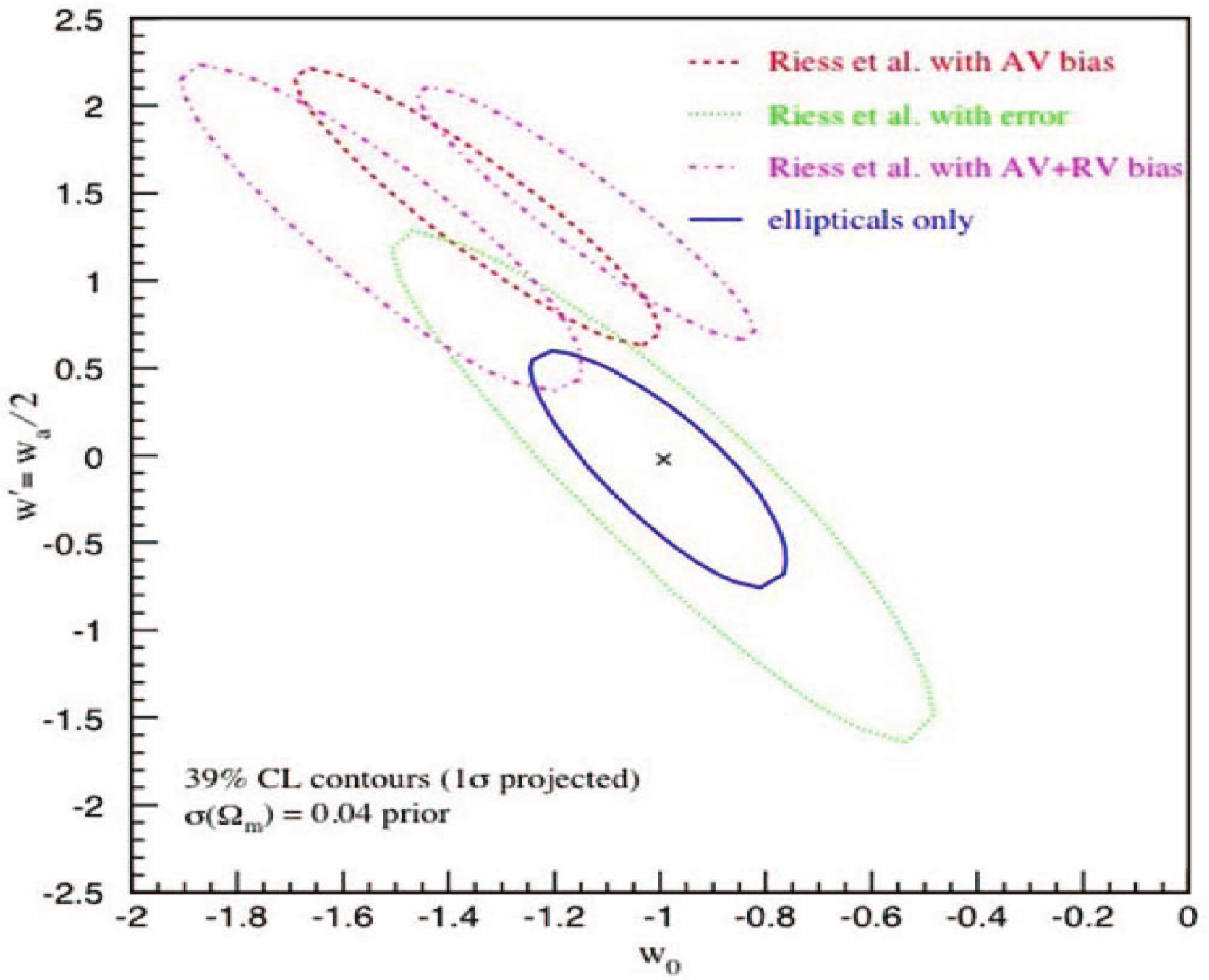}
\caption{\small Example of the increase of errors due to dust extinction correction, and of biases
due to uncorrected extinction.
\label{fig:dustBias}}
\end{center}
\end{figure}

Alternatively, surveys can restrict themselves to
SNe with low extinction, signaled either by their
low measured values of $E(B-V)$ or by its location in an old elliptical galaxy where star
formation has long ceased and dust presence is minimal. Figure~\ref{fig:dustBias} shows a
$w_0-w_a$ contour plane with
the qualitative effect of dust correction through measurement of $A_V$ and $R_V$ from data
(which increases the contour size significantly), and of uncorrected dust biases (which displace
the contour).

``Gray'' dust, with an effective $R_V \to \infty$, had been postulated as an explanation for
the observed dimming of SNe at large redshift. The correction method outlined above would not
work for a dust that would dim equally all wavelengths. However, natural models of gray dust
would lead to dimming of all SNe at all redshifts. This has been excluded 
by~\cite{knop03,riess04}, which have observed SNe at redshifts beyond $z=1$ and found them to be
brighter, not dimmer, than expected by models without dark energy, and in perfect agreement
with the prediction of the ``concordance'' model: 
$\Omega_M \approx 0.25$, $\Omega_\Lambda \approx 0.75$.
\subsection{Flux calibration}\label{sec:calib}
By flux calibration we understand the determination of the zero points $\phi_{0,j}$ for each
filter $j$. While the overall normalization is irrelevant (since it can be absorbed in the
unknown parameter ${\cal M}$ in eq.~(\ref{eq:hubble})), the relative filter-to-filter 
normalizations are crucial, as they influence, for instance, the determination of colors,
which are needed for the dust-extinction corrections (as we saw in the previous section),
K-corrections, etc.

\begin{figure}[tb]
\begin{center}
\includegraphics[scale=0.65]{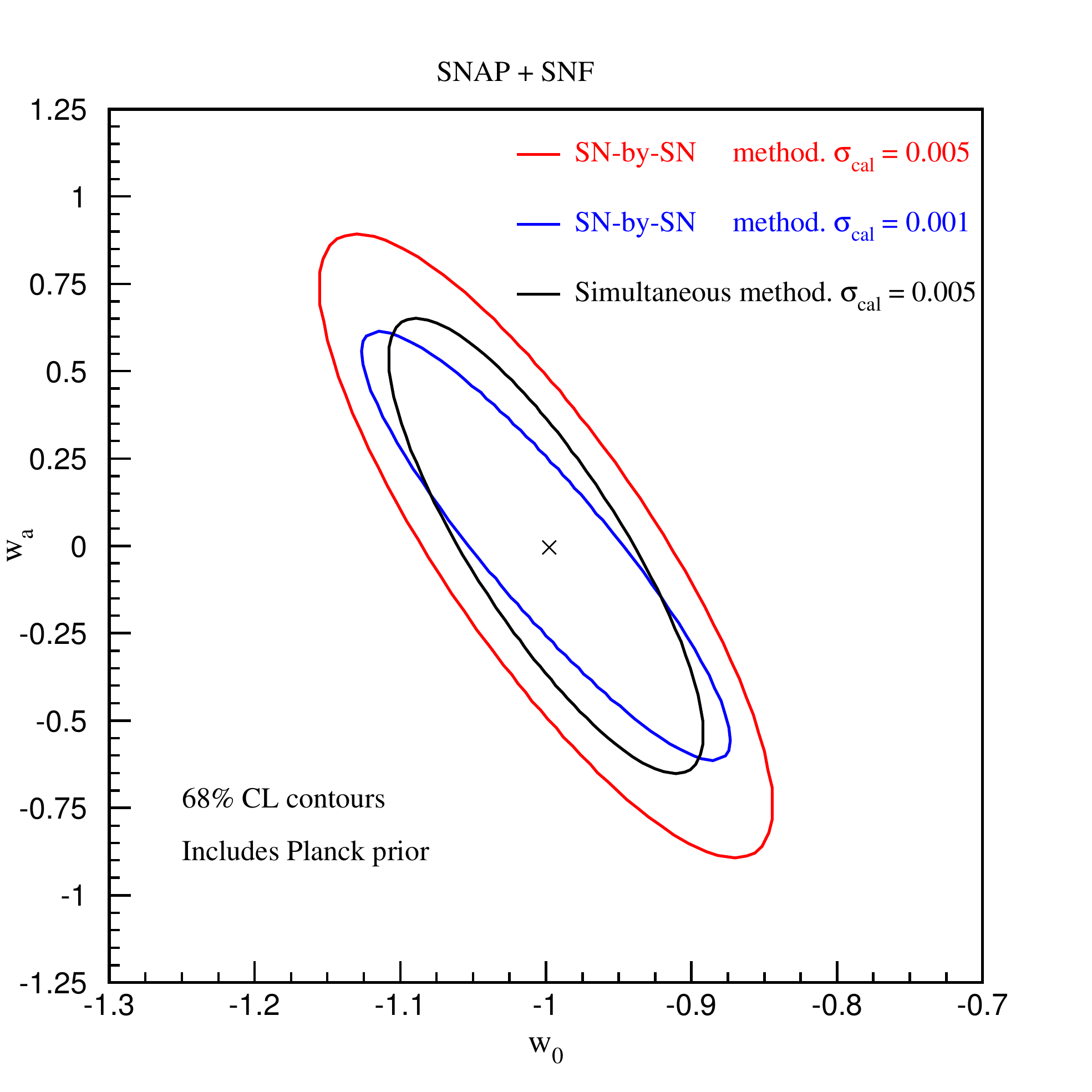}
\caption{\small Effect of self-calibration in a survey similar to the one proposed by the SNAP
collaboration. Using the procedure in~\cite{KM} and assuming an external calibration error
of 0.005 is roughly equivalent to using the standard procedure with an external error of 0.001.
\label{fig:cali}}
\end{center}
\end{figure}

The standard procedures use well-understood stars or laboratory light sources to achieve values
of $\sigma_{\mathrm cal}$ around few per cent in flux. A complementary procedure has been
presented in~\cite{KM} which uses supernova data themselves to achieve a large degree of 
self-calibration. For example, fig.~\ref{fig:cali} show that for a fiducial survey close to the
SNAP mission specifications, the procedure of~\cite{KM} achieves an effective factor 5 reduction 
in calibration error.
\section{Current surveys: SNLS}\label{sec:snls}
\begin{figure}[tb]
\begin{center}
\includegraphics[scale=0.75]{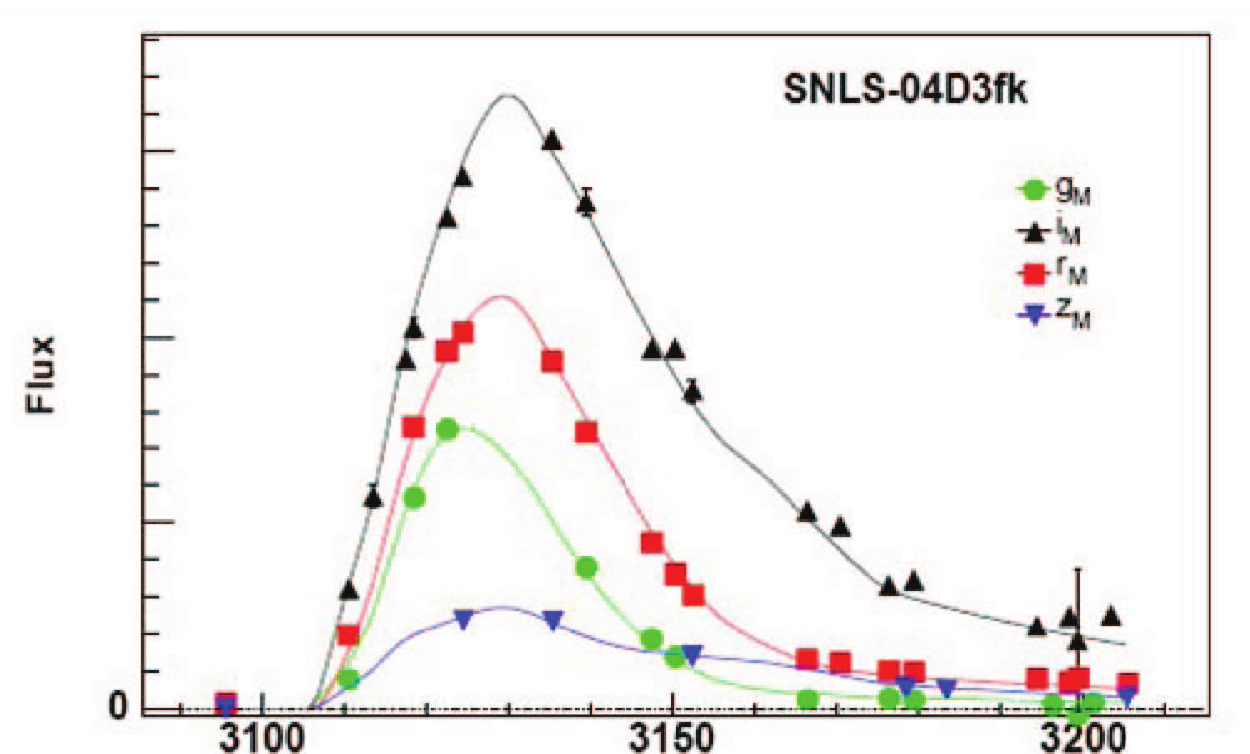}
\includegraphics[scale=0.75]{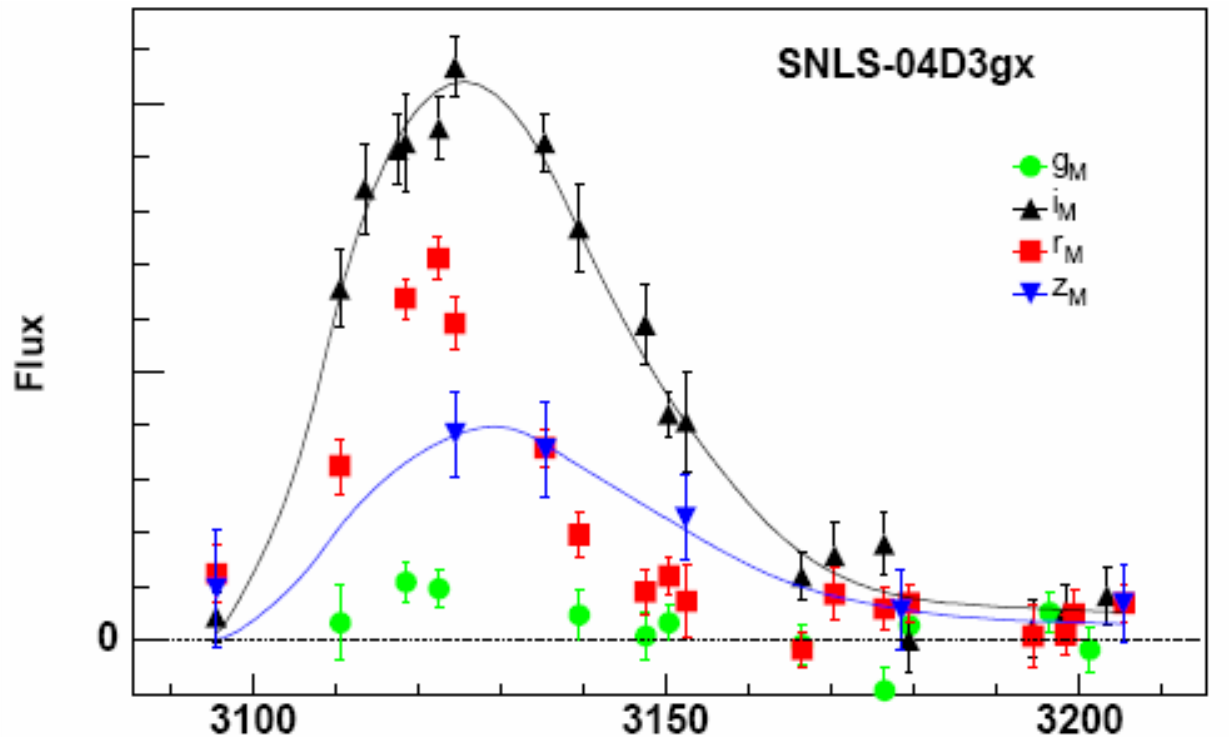}
\caption{\small SNLS light curves for low- (top), and high- (bottom) redshift supernovae. 
For the SN at redshift $z=0.91$ the $g$ and $r$ light-curves correspond to deep UV in the
SN rest frame and, therefore, are not used in the analysis. The horizontal scales are in units
of days. 
\label{fig:snls-data}}
\end{center}
\end{figure}

Of the current type-Ia SN surveys, the most promising is probably the SuperNova Legacy Survey
(SNLS)~\cite{snls}, taking place at the Canadian-French-Hawaiian Telescope (CFHT) in the
Mauna Kea observatory in Hawai'i. The telescope
and camera provide a 3.6~m aperture, 1 deg$^2$ of field of view and a focal plane with 36 CCDs
with a total of 328 million pixels.
The survey team will be taking data for 40 nights per year 
during the 2003-2008 period in a four-night-cadence 
rolling search in four 1-deg$^2$ fields in the $g$, $r$, $i$, and $z$ bands. At the end
of the five years the collaboration expects to have discovered and followed 
500-700 type-Ia SNe with
redshifts up to $z=1$. Spectroscopic follow-up of most good candidates is performed in
several 10~m-class telescopes in both hemispheres: VLT, Gemini North and South, and Keck. The
resulting spectroscopic time needed is comparable in size to the imaging time in CFHT.
Figure~\ref{fig:snls-data} shows two SNe with the four light curves measured for each one.
For the SN at redshift $z=0.91$ the $g$ and $r$ light-curves correspond to deep UV in the
SN rest frame and, therefore, are not used in the analysis.
For each SN $i$, the SNLS light-curve fit performs K-corrections and returns the $B$ magnitude 
(in SN rest frame) at peak luminosity $m_B^i$, the stretch factor $s^i$, and the observed color
excess $E^i(B-V)$. Every available filter is used in the fit, provided it corresponds to the
$U$, $B$, $V$, $R$ regions in the SN rest frame. At least two filters for SN are required (in
order to determine $E(B-V)$. The cosmology fit then proceeds as:
\begin{equation}
m_B^i = {\cal M}_B + 5\log [H_0 d_L(z^i, \vec{\theta})] - \alpha (s^i-1)
+\beta E^i(B-V) \, ,
\end{equation}
where $\vec{\theta}$ are the cosmological parameters, and the
nuisance parameters ${\cal M}_B$, $\alpha$, $\beta$ are also fitted from the data.
The last two give the slopes of the
dependency of the magnitude with stretch ($\alpha$) and color ($\beta$).

\begin{figure}[tb]
\begin{center}
\includegraphics[scale=0.65]{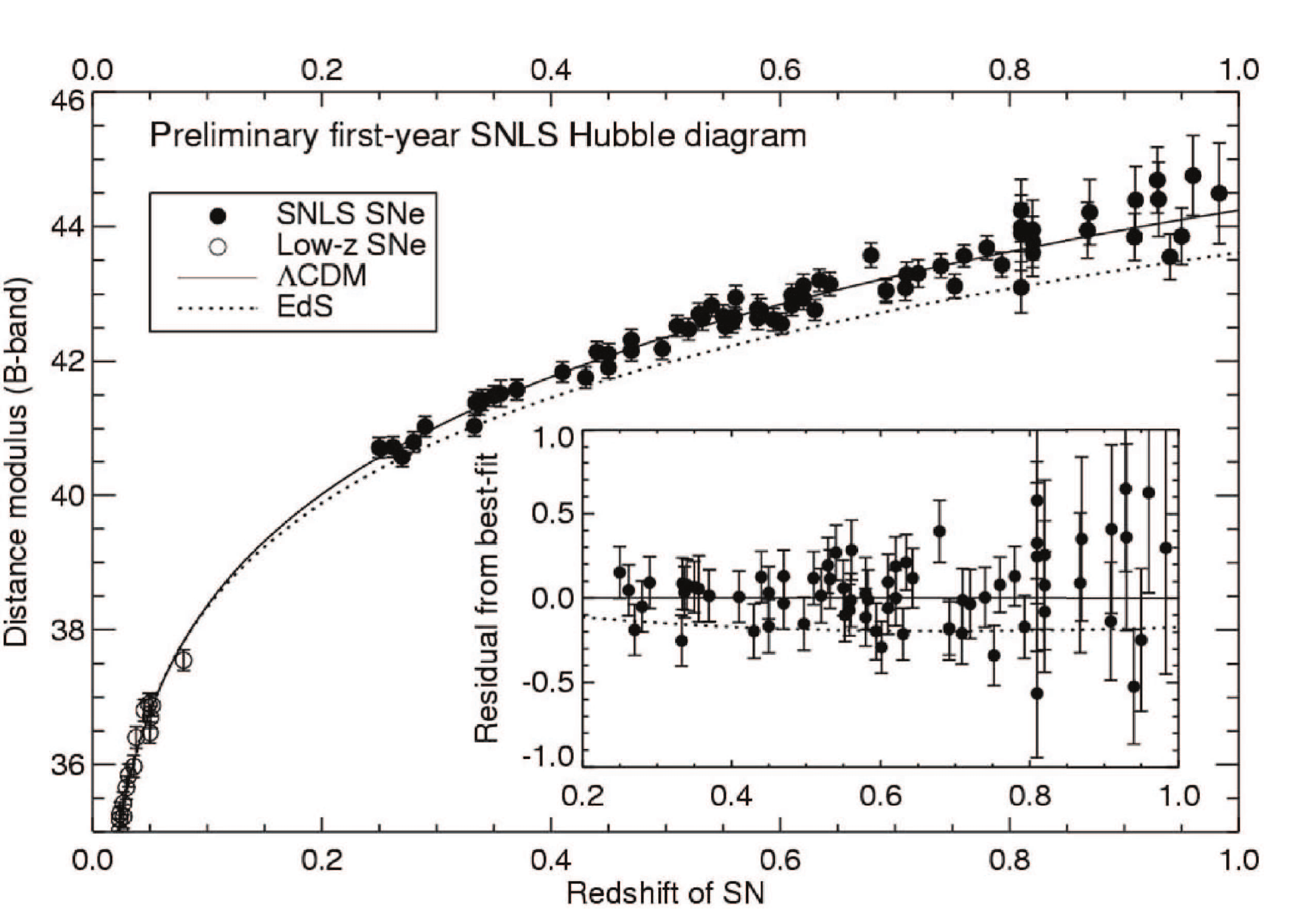}
\caption{\small Hubble diagram corresponding to the SNLS first-year data, together with 
44 low-$z$ SNe. Note that the average dispersion of the SNLS
measured magnitudes about the best-fit prediction is only 0.12.
\label{fig:snls-hubble}}
\end{center}
\end{figure}

Analyzing 73 new high-$z$ SNe from SNLS first-year data, together with 44 previously published
near-by ($z < 0.1$) SNe, and including the Baryon Acoustic Oscillations measurement 
of~\cite{eisenstein}, the SNLS team finds 68\% constraints on the cosmological parameters:
\begin{eqnarray}
\Omega_M = 0.271 \pm 0.022  \nonumber \\
w = -1.02 \pm 0.11 \, ,
\end{eqnarray} 
where $w$ has been assumed constant and the universe flat. Figure~\ref{fig:snls-hubble} shows
the Hubble diagram corresponding to these data. The mean dispersion of the SNLS
magnitudes about the
best-fit prediction is only $\sigma_{\mathrm{int}} = 0.12$~mag.

The SNLS also expects to gather an additional 500 or so type-Ia SNe without spectroscopic data,
because of lack of resources. This highlights the problem that will be faced by the next
generation of deep ground-based type-Ia SN surveys. 
For instance, the Dark Energy Survey (DES) plans
to image about 2000 type-Ia SNe up to $z=1$ in 2010-2014. However, it seems impossible 
to gather the 10~m-class telescope needed to get spectra for all those SNe. Therefore, techniques
have to be developed to do without spectroscopy for most of the SNe. A recent paper by
SNLS~\cite{sullivan} discusses some methods to classify SNe with just multi-color light-curve
information, and to determine their photometric redshift. A 90\% purity is obtained by using a 
real-time analysis of pre-maximum light curves (used to trigger spectroscopic follow-up).
Presumably, higher purity and efficiency could be achieved by using all light-curve information.
Similarly, the precision of the photometric redshifts is around 
$\langle|z_{\mathrm phot} - z_{\mathrm spec}|\rangle = 0.03\cdot(1+z)$, 
which seems adequate for most purposes.
\section{Future surveys: SNAP}\label{sec:snap}
Of the proposed supernova surveys in the next decade, the SuperNova Acceleration Probe (SNAP) 
satellite mission is probably the most ambitious. In essence, it consists of a 2~m-class 
wide-field (0.7~deg$^2$) imager with state-of-the-art optical and near-infrared camera and
an integral-field-unit spectrograph. The dual aim is to collect about 2000 type-Ia SNe up to
redshift $z = 1.7$, and to study weak gravitational lensing from space. If approved, it could
fly from about 2013--2015 on.

\begin{figure}[tb]
\begin{center}
\includegraphics[scale=0.65]{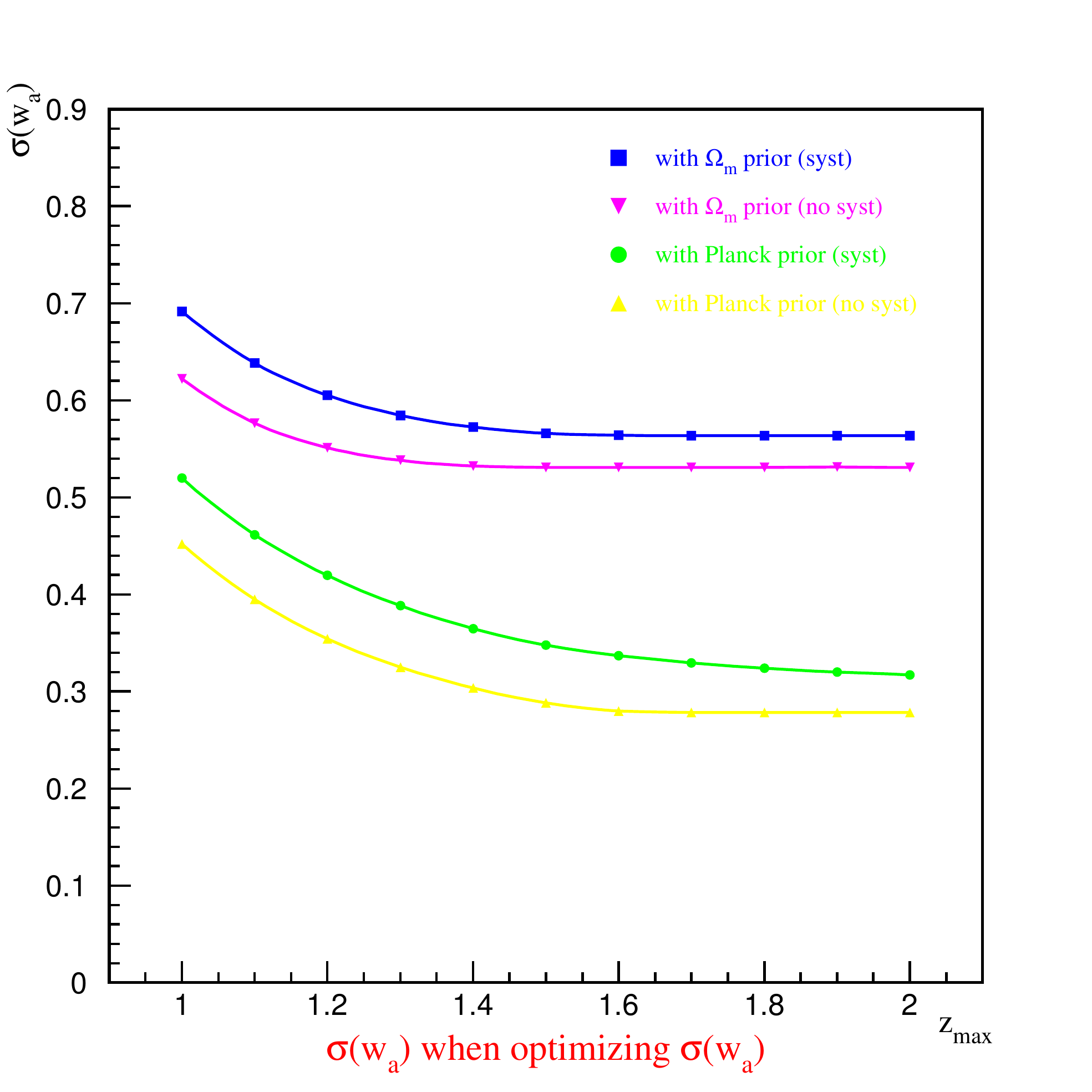}
\caption{\small Uncertainty on the $w_a$ parameter as a function of maximum redshift 
$z_{\mathrm max}$ for a SNAP-like mission with fixed total time. The different lines correspond
to different assumptions about priors and systematic errors.
\label{fig:zrange}}
\end{center}
\end{figure}

Since a space mission is always much more expensive than a ground-based survey, the first question
that comes to mind is ``why space''? Figure~\ref{fig:zrange} demonstrates that for a SNAP-like
mission, and keeping the time of the mission constant, there is a clear advantage in sensitivity
to $w_a$ by going to larger redshifts, $z\geq 1.5$. Furthermore, the window to the deceleration
era $z>1$ can help in eliminating systematic errors (see, for instance, the gray dust discussion in
section~\ref{sec:dust}). However, for $z>1-1.2$, the rest-frame $B$ band gets redshifted into the
observer near infra-red region ($\lambda > 1.2 \mu$m). At these wavelengths, atmospheric 
absorption makes it all but impossible to perform accurate measurements from the ground, hence
the need for a space-based mission.

The SNAP optical imaging system incorporates nine redshifted wide-band filters covering from the
$U$ band to about 1.7~microns. The detectors are LBNL-developed thick, back-illuminated
CCDs with quantum efficiency above 50\% up to one~micron, and HgCdTe detectors covering the
near-infrared region. The nine filters ensure that at least there colors are available for
all SNe in their restframe $U$ to $R$ wavelength range. This information can be used
to determine the reddening law $R_V$ individually for each supernova, eliminating a potentially
damaging systematic uncertainty.

The large number of SNe in each redshift bin will also allow to tackle the issue of ``evolution''
of SNe properties with redshift. This has been often mentioned as a potentially dangerous source
of systematic errors. By taking multi-color light-curves and multi-epoch spectra for all the
supernovae, SNAP will be able to classify them according to their observational differences.
Then, cosmology can be extracted by performing cosmology fits within each sub-type, each
including both low- and high-redshift SNe (``like-to-like'' comparison). In practice,
this is done by allowing a different value of the nuisance parameter ${\cal M}$ for each
sub-type. It has been shown in~\cite{KLMM} that the statistical degradation due to the extra
free parameters in the fit is only of a few per cent.

\begin{figure}[tb]
\begin{center}
\includegraphics[scale=0.85]{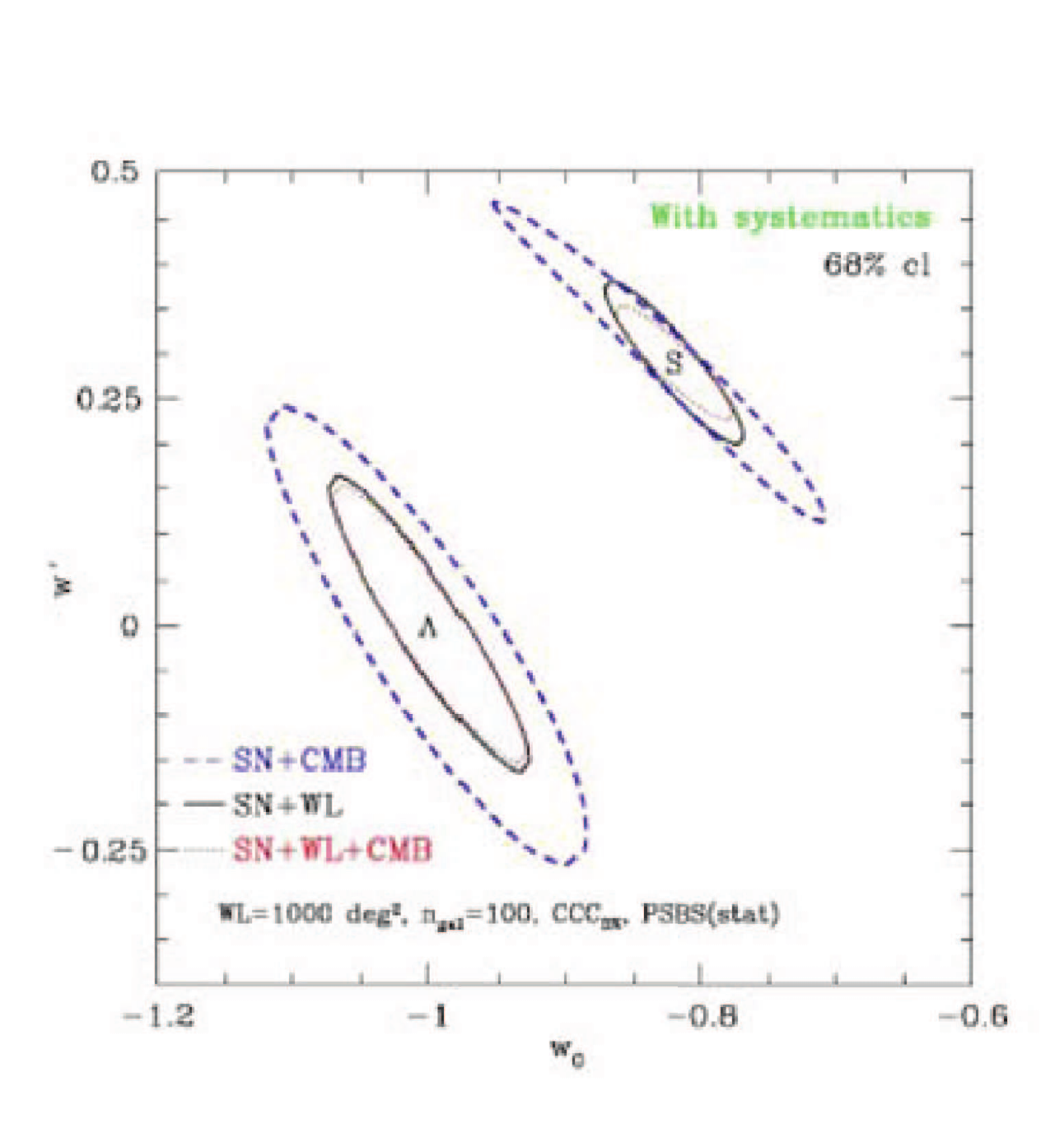}
\caption{\small Expected reach of the SNAP satellite mission. The ``$\Lambda$'' contours
correspond to assuming a $\Lambda$CDM fiducial universe, while the ``S'' contours correspond 
to a Supergravity-inspired model. The ``$\Lambda$'' universe tends to lead to the most
conservative contours. Note in both cases the big improvement after
adding weak lensing.
\label{fig:snap}}
\end{center}
\end{figure}

Figure~\ref{fig:snap} shows the expected SNAP 
precision in the $w_0$--$w_a$ plane. SNAP with SNe and
weak lensing can, by itself, determine $w_0$ to about 5\%, and $w'\approx w_a/2$ to about 10\%.
\section{Summary}\label{sec:summary}
Type-Ia supernovae provided the ``smoking gun'' for the accelerated expansion of the universe
and the existence of dark energy. It is by now a mature technique, where sources of
systematic errors are better identified and understood than for most other techniques. 
However, it is still being
perfected, and improvements happen constantly.

The control of systematic errors holds the key to any substantial future improvements. 
Calibration, redshift evolution of SN properties, and dust extinction corrections are the three
main sources of systematic uncertainties. Current and future SN surveys address them with 
different levels of sophistication. 

There is a vigorous current and future program of SN surveys, ranging from low-$z$ SNe from
the ground (like SNF, SDSS-II/SNe, CfA, Carnegie) through medium- to high-$z$ surveys from
the ground (Essence, SNLS, DES, PanSTARRS, LSST) to high-$z$ surveys from space
(JDEM, DUNE). 

We should expect more insight on the nature of dark energy from current and future studies of
type-Ia supernova samples.
\section*{Acknowledgments}
It is a pleasure to thank the organizers of the conference and in particular Joan Sol\`a for
their kind invitation and for running the conference so smoothly.
\section*{References}
\end{document}